\documentclass[twoside]{dis09}
\usepackage[latin1]{inputenc}
\usepackage[dvips]{graphicx,epsfig,color}
\usepackage{wrapfig,rotating}
\usepackage{amssymb,amsmath,array}

\begin{document}
\title{First Measurements with the LHCb Experiment}

\author{Markward Britsch on behalf of the LHCb collaboration
%
%
\vspace{.3cm}\\
%
Max-Planck-Institut f\"ur Kernphysik \\
PO Box 103980, 69029 Heidelberg - Germany
%
}

\maketitle

\begin{abstract}
The LHCb detector covers the forward angular region 
complementary to the angular coverage by the
general purpose LHC experiments. With the first data, measurements of inclusive
charged particle
production and $V^0$ production will be performed and compared to particle
production models. Focusing on ratios, such as positives/negatives or
lambda-bar/lambda, many systematic effects
cancel and publishable results are expected soon after the turn-on of LHC. The
talk will present the analysis strategies, the physics questions and the
results expected from the first $10^8$ minimum bias $pp$-collisions at a
center-of-mass energy of 10 TeV, which can be recorded within one day of stable
running.

\end{abstract}

\section{Introduction}
\label{s:intro}

 \footnote{For the slides of this talk see~\cite{url}.}
The baryon asymmetry of the universe indicates that there should be a source of
CP violation beyond that from the Standard Model. Alternative models often predict the existence of new kinds of
particles. One way to tackle this puzzle are precision measurements of
$CP$ violation and rare decays which are in general sensitive to new kinds of
heavy particles from loop diagrams. 
LHCb~\cite{LHCb} is an experiment at the $pp$-collider LHC
dedicated to these kinds of measurements. This means that we are mainly
interested in heavy flavor physics. As most $b\bar{b}$-quark pairs are produced
in forward and backward direction, LHCb was build as a forward spectrometer
with a pseudorapidity coverage for $B$-mesons of $1.9 < \eta < 4.9$. This is
complementary to the central $\eta$-coverage of ATLAS and CMS of about $-2.5 <
\eta < 2.5$. In addition LHCb can measure down to lower transverse momenta.

To achieve its ambitious goals, the LHCb experiment has a good vertex
resolution, dedicated triggers and various precise particle identification
(PID) systems including two Ring Imaging Cherenkov counters (RICH). 
In the following we will concentrate on the
tracking system which consists of the vertex detector, called Vertex Locator or
VeLo, two tracking stations in front of the magnet (called TT stations) and the
main tracker consisting of three stations behind the magnet. The main tracker is
split into the inner and outer part. The inner tracker is much smaller than
the outer tracker and covers the largest pseudorapidities where the track
multiplicities are highest.


The conditions of the LHC in 2009/2010 will be different from the nominal ones
with a lower center-of-mass energy of $\sqrt{s} = 8-10$~TeV and a lower
luminosity of up to ${\cal L} \sim 10^{32}$~cm$^{-2}$s$^{-1}$. The nominal
values for ${\cal L}$ and $\sqrt{s}$ are not the crucial parameters for the LHCb experiment and
its full physics program can start in 2009/2010.
The LHCb experiment's commissioning is well on the way using cosmic data as well
as beam--gas and beam on collimator data from the LHC commissioning in fall
2008~\cite{deCapua}. 


First measurements will be done using events collected by a minimum bias
trigger and the nominal data logging rate of LHCb of 2 kHz. 
With this we can expect to record
$10^{8}$ events within the first few days after the start of LHC. 
Initially, we will only use tracking without PID as long as the PID system is not
calibrated. The first goal will be to measure particle ratios like those for
charged tracks, $K^0_s$, $\Lambda$ or $D$-mesons, i.e., particle to
anti-particle ratios as well as $K^0_s$/$\Lambda$. Here most systematics cancel
and no luminosity measurement is needed in contrast to cross section
measurements. The studies discussed below are based on $9.5\cdot 10^{6}$ LHCb Monte Carlo
events produced in 2006 with $\sqrt{s} = 14$ TeV. 

\section{Physics topics}

The physics topics covered here are inclusive production, strangeness production
and charm signals, all shown to be feasible with the very first data. While
interesting in their own right, they will serve as stepping stones to heavy flavor physics,
especially to $B$-decays with $K^0_s$ as daughter, $b$-baryon spectroscopy and lifetimes, and radiative b-baryon decays. In addition
they will be used as input for tuning the Monte Carlo event
generator and used to test
fragmentation models for multi particle production.

\begin{wrapfigure}{r}{0.55\columnwidth}
\centerline{\includegraphics[width=0.55\textwidth]{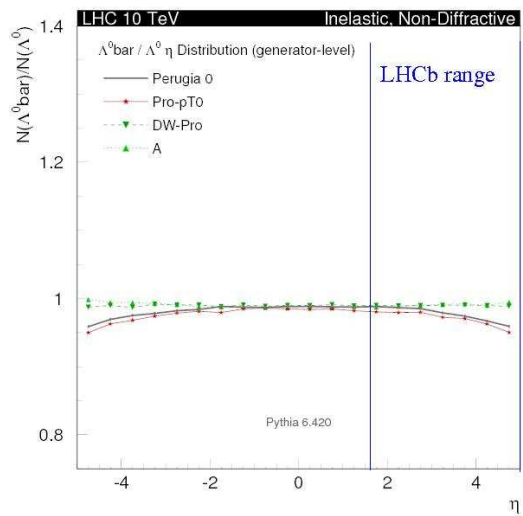}}
\caption{$\bar{\Lambda}$ to $\Lambda$ number ratio plotted versus pseudorapidity
  for old (triangles) and new (no symbol, star) models at LHC. The vertical lines mark the LHCb range. Taken from~\cite{skands}.}\label{f:eta}
\end{wrapfigure}

Elements of multi particle production 
are fragmentation, color (re)con\-nec\-tion
and multiple parton interaction (MPI) which will be especially important at the
LHC. Color (re)connection deals with the multitude of possibilities to connect
the differently colored lines of the quarks and gluons in the corresponding
Feynman diagram to result in a color neutral state in the end. MPI means that there is
more than one parton of each proton involved in the collision. In recent years
new models have been developed, see, e.g.,~\cite{sjostrand}. 

Strangeness is an additional probe for the fragmentation process, since the $s$-quark
is still a comparatively light particle, it is mainly created in the
hadronization stage. 
In addition some
new models predict the beam baryon number to reach lower $\eta$ at low
$p_t$~\cite{sjostrand}. 
This can be seen in Figure~\ref{f:eta} where the ratio of
$\bar{\Lambda}$ to $\Lambda$ 
are plotted versus
pseudorapidity for two old and two new models. In the central pseudorapidity range the
models hardly differ while in the range of LHCb indicated in the figure shows
that the difference is up to 5~\%.

\section{Inclusive production}

For the charged track ratios we generated one million minimum bias events and
selected tracks in the range $100$ MeV/$c$ $< p_t < 8000$ MeV/$c$ and $1.8 <
\eta < 5.1$. A minimal requirement for this measurement to be possible will be a
working main tracker. These kind of studies will be vital for understanding charge
asymmetries but will also be used for Monte Carlo tuning and comparison with
different fragmentation models. 


\begin{figure}
\centerline{\includegraphics[width=0.45\textwidth]{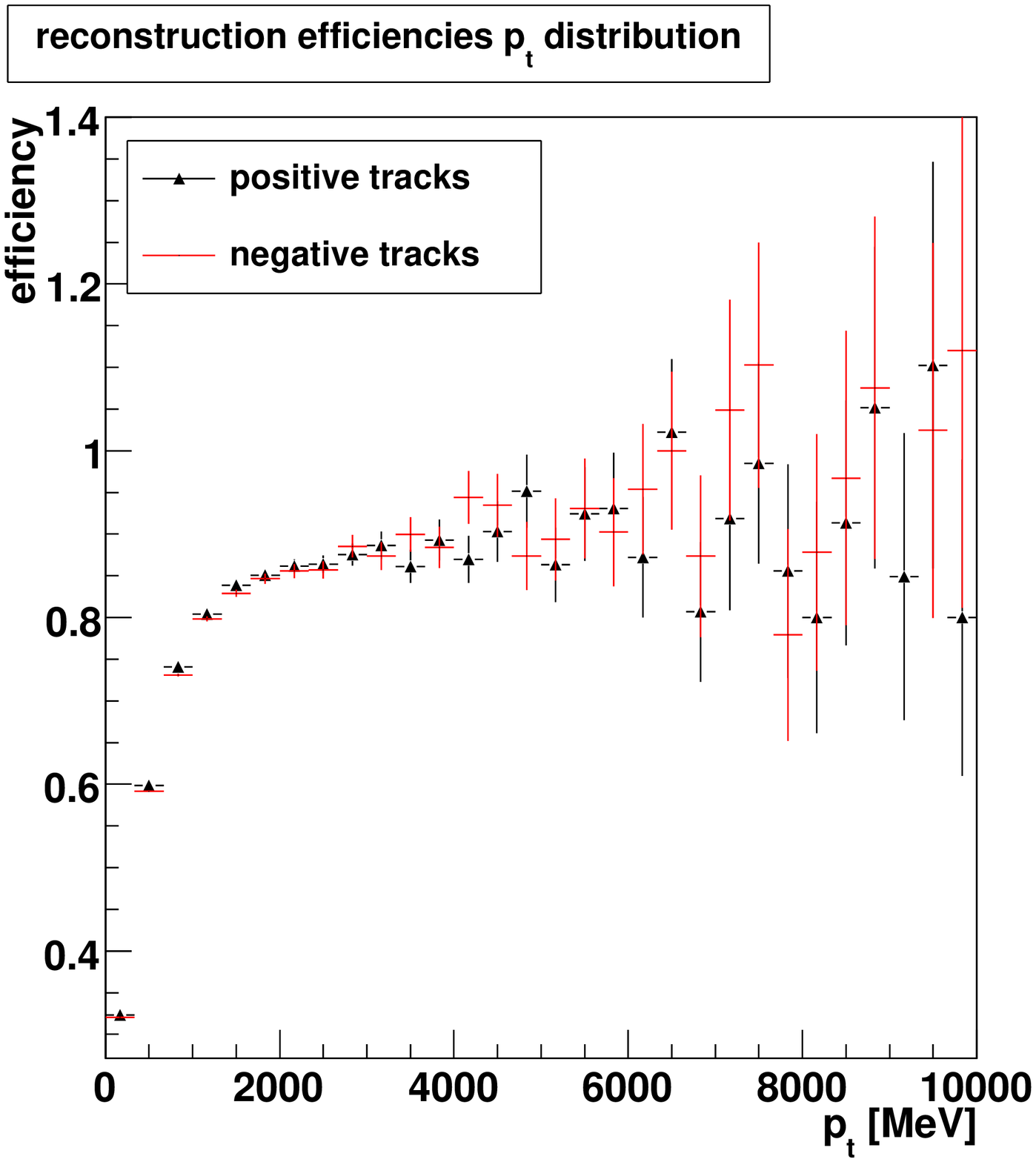}\includegraphics[width=0.45\textwidth]{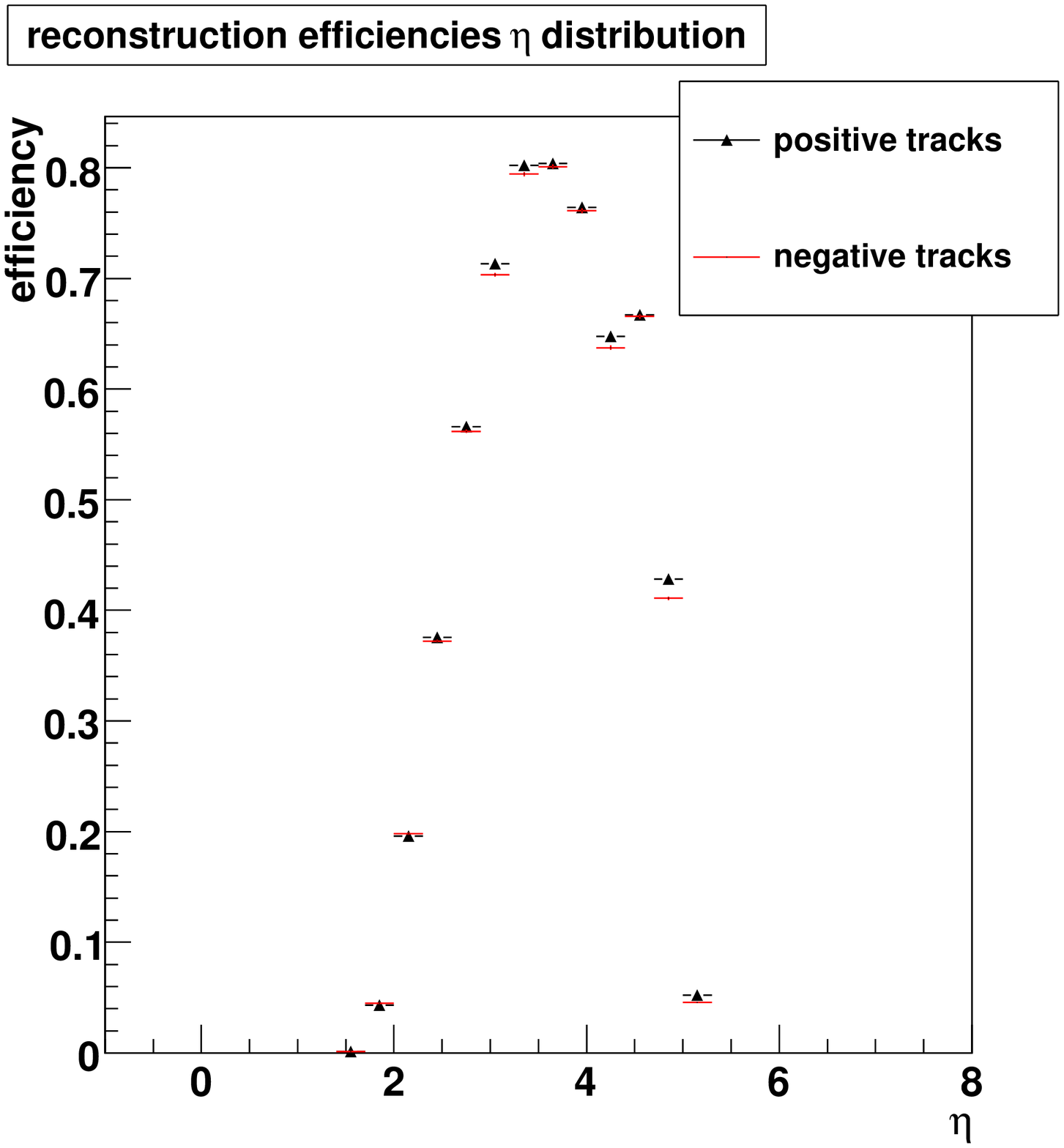}}
\caption{Reconstruction efficiencies for positive (black, triangles) and negative
  (red) tracks plotted versus transverse momentum (left) and
  pseudorapidity (right). For the $p_t$ distributions particles are integrated over $1.8 <
  \eta < 5.1$, for the $\eta$-distributions over 100 MeV/$c$ $< p_t < 8000$ MeV/$c$. }\label{f:eff}
\end{figure}

\begin{figure}
\centerline{\includegraphics[width=0.45\textwidth]{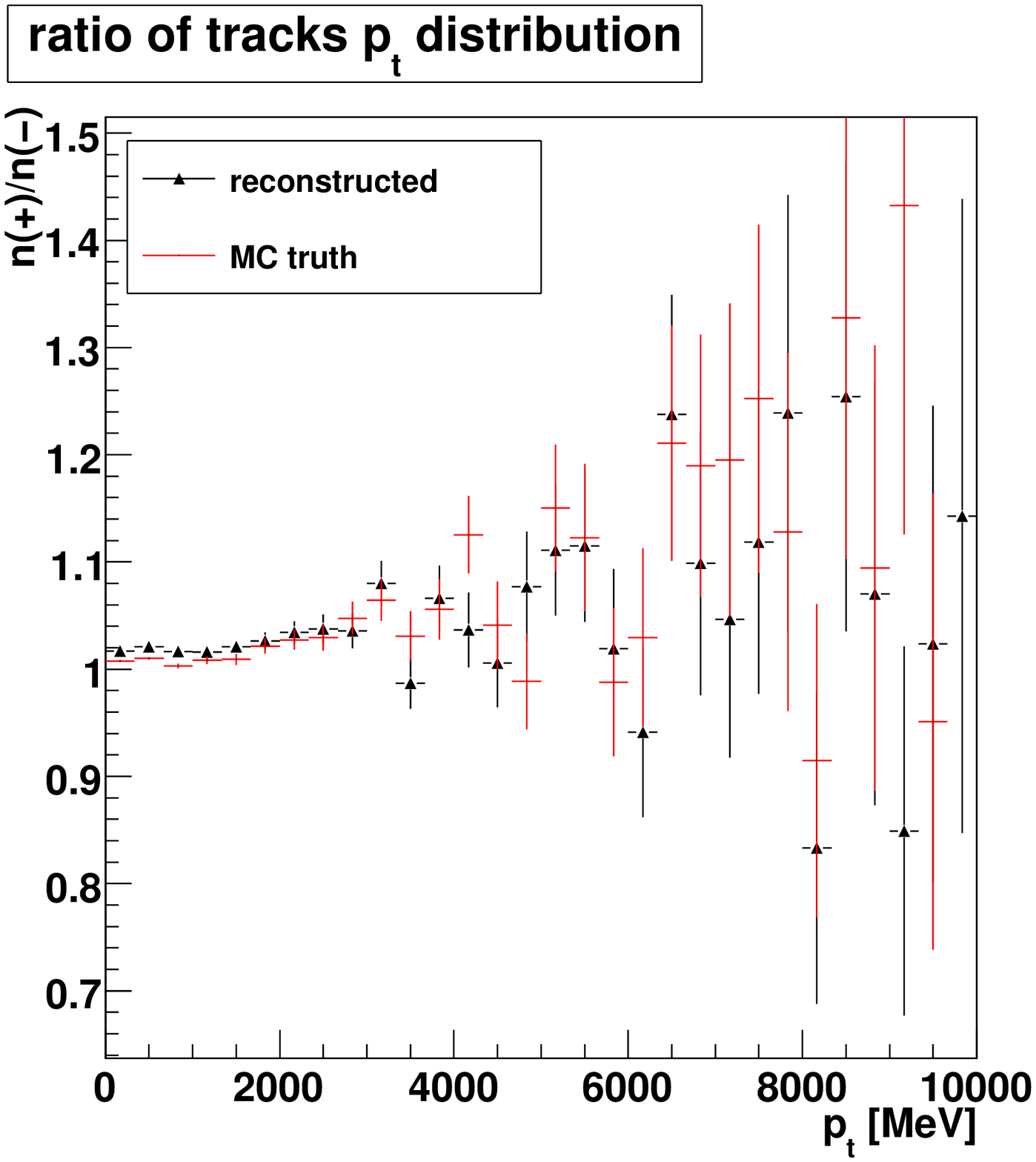}\includegraphics[width=0.45\textwidth]{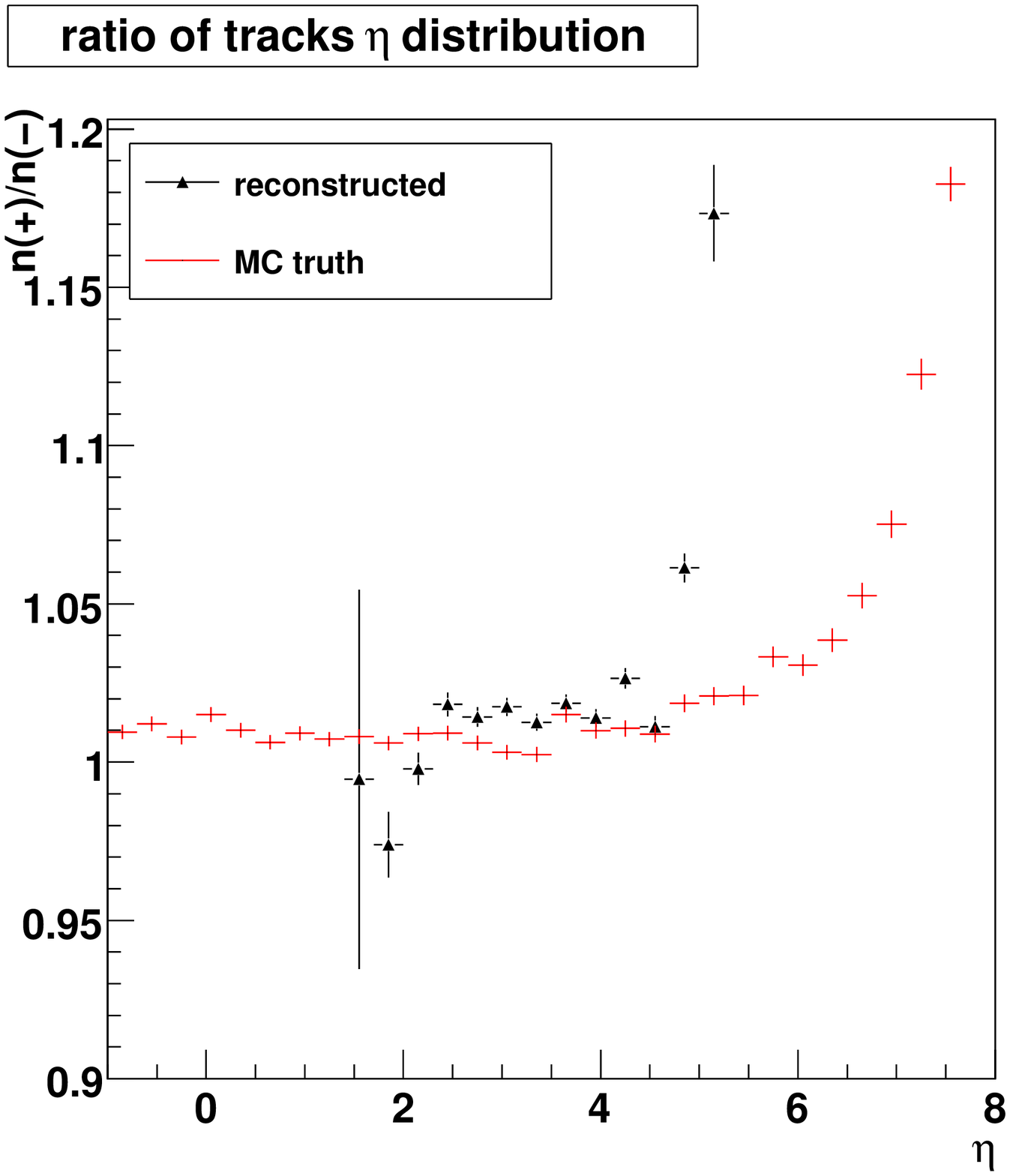}}
\caption{Number ratio of positive to negative tracks as reconstructed (black, triangles)
  and generated (red) plotted versus transverse momentum (left) and
  pseudorapidity (right).}\label{f:chargeRatio}
\end{figure}

To compare the reconstructed tracks to the simulated ones we used the
following selection criteria. For the generated tracks, only those from the
events with
exactly one primary vertex (PV) 
have been accepted. Both elastic and diffractive events were rejected. Only long lived particles (lifetime $\tau > 1$ ns) in our
$(\eta,p_t)$-range coming
from the PV or from a short lived particle from the PV have been selected.
For the reconstructed tracks only events with exactly one PV have been used. The
track selection required the tracks to be compatible with coming from the PV
(i.e., impact parameter $<$
0.15 mm) and to have hits in both VeLo and main tracker.

In Figure~\ref{f:eff} the reconstruction efficiencies versus transverse momentum
($p_t$) and pseudorapidity are shown. 
Figure~\ref{f:chargeRatio} shows the charged track ratios for generated and
reconstructed tracks as a function of $p_t$ and $\eta$. The differences
between generated and reconstructed ratios are $\lesssim$ 5~\% for most of the
bins. 
Still, it will be necessary to correct the reconstructed ratios for acceptance. 
If we use the correction factors obtained from the simulation, we need to trust the simulation result to a level of 20~\% in order to
get a systematic error of $\sim$ 5~\%$\times$20~\% = 1~\%.

\section[Identified particles]{Cross section ratios for identified particles}

For the strange particle selection we use the decays $K^0_s \rightarrow \pi^+
\pi^-$, $\Lambda \rightarrow p \pi^-$ and  $\bar{\Lambda} \rightarrow \bar{p}
\pi^+$. Candidates are pairs of oppositely charged tracks. 
The selection considered in this study is based on 
the values of a reconstructed variables rather than their significances. 
\begin{wrapfigure}{r}{0.4\columnwidth}
\centerline{\includegraphics[width=0.38\textwidth]{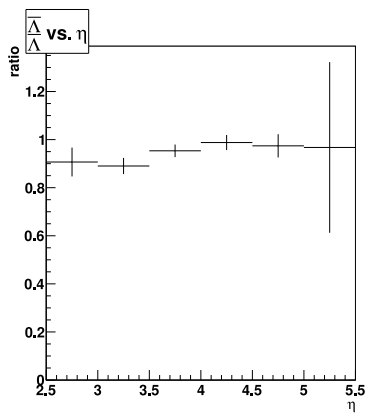}}
\caption{Reconstructed $\bar{\Lambda}$ to $\Lambda$ number ratio plotted versus pseudorapidity.}\label{f:lamRatio}
\end{wrapfigure}

Although the later will improve the sensitivity of the analysis,
they will only be employed at a later stage since they require a more complete
understanding of the detector. The studies are based on 9.5 million minimum bias
events, i.e., an order of magnitude less than what we expect to
have as the first data (see Section~\ref{s:intro}). 
Minimal requirements for these measurements
will be a working VeLo and main tracker. It can be used to
check momentum calibration and is important for RICH calibration. 

\begin{figure}
\centerline{\includegraphics[width=0.32\textwidth]{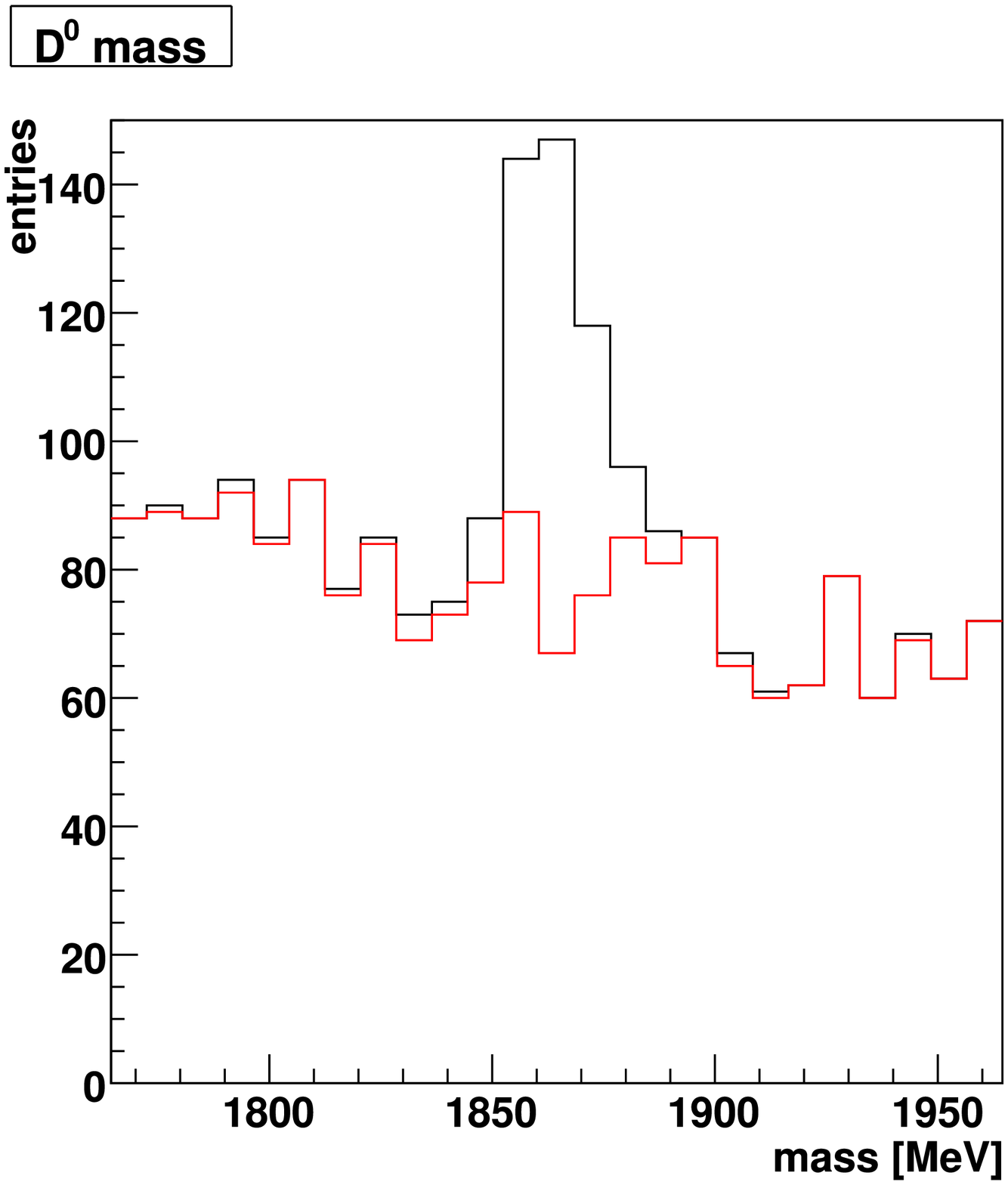}\hspace{0.00\textwidth}\includegraphics[width=0.32\textwidth]{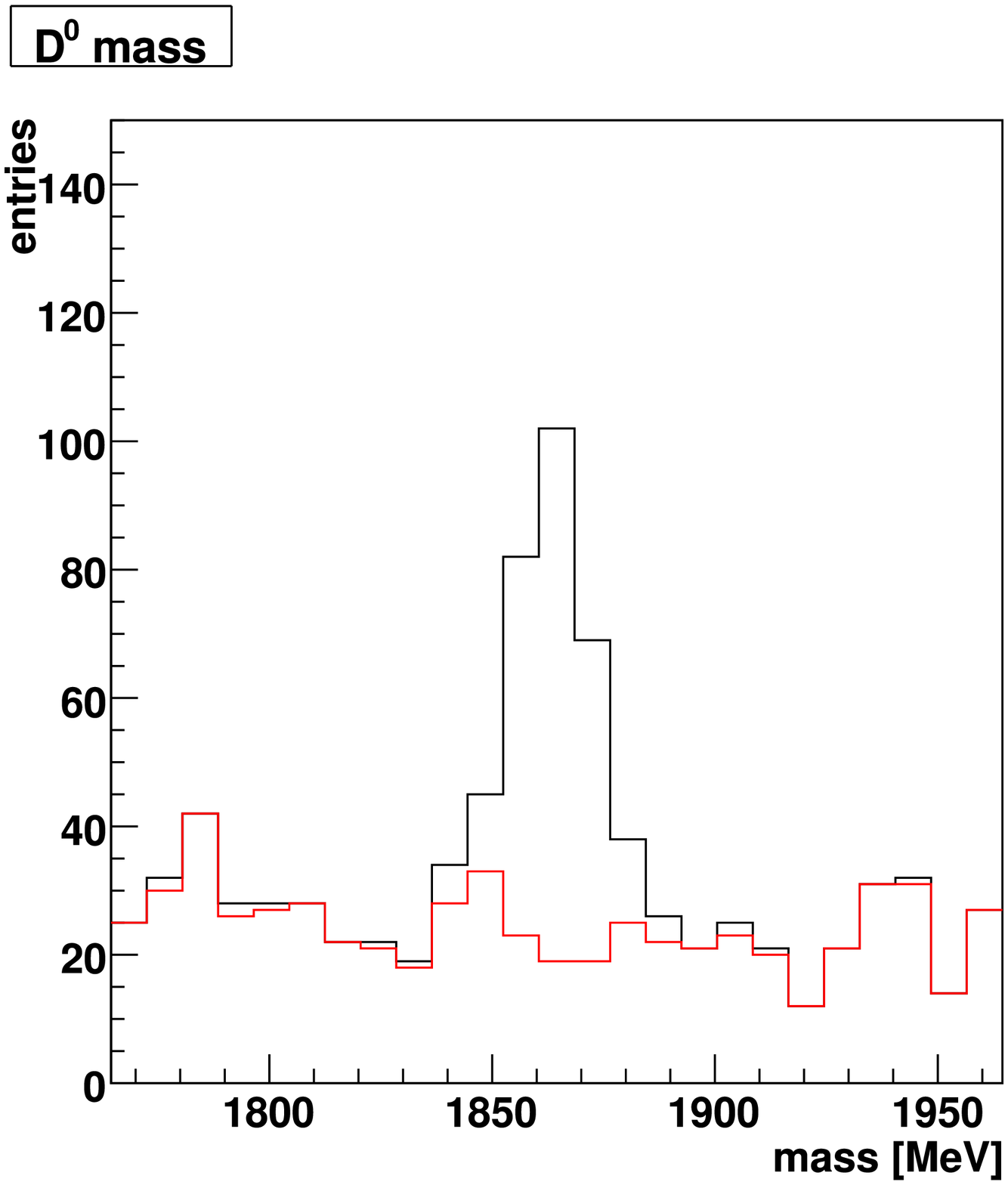}\hspace{0.03\textwidth}\includegraphics[width=0.32\textwidth]{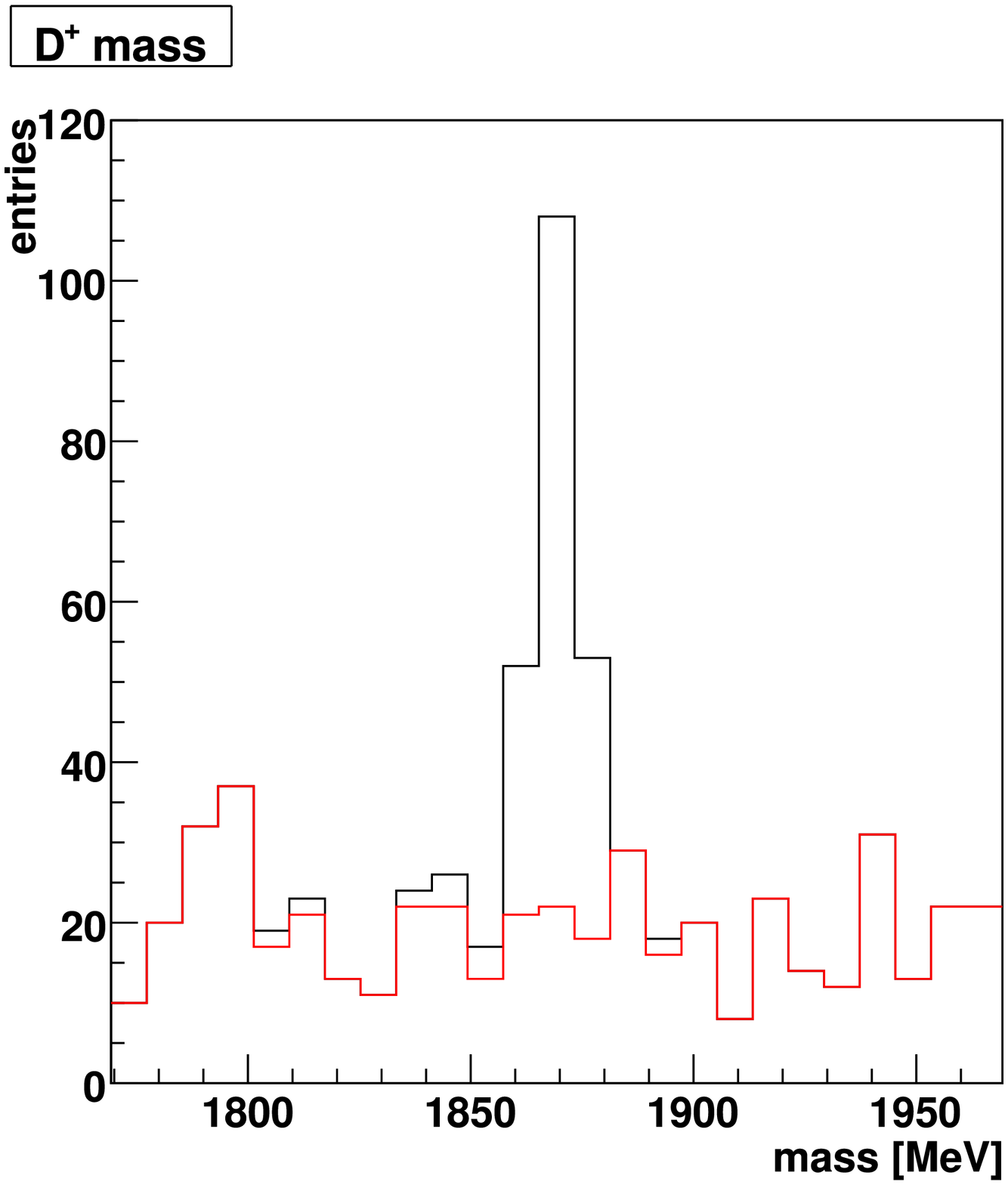}}
\caption{$D^0$ mass peak for the cuts based (left) and MVA (center)
  selection. The MVA parameters have been chosen in a way to get the same signal
  yield for comparison. On the right the $D^+$ mass peak using the cuts based
  selection is shown.}\label{f:D0mass}
\end{figure}

Figure~\ref{f:lamRatio} shows the reconstructed ratios of $\bar{\Lambda}$ to
$\Lambda$ as a function of pseudorapidity. $\Lambda$ and $\bar{\Lambda}$ were
selected requiring that the 
distance of closest approach between the two tracks
$\le$~0.3~mm, flight-length$\cdot \gamma \beta$ $\ge$~4~mm, impact parameter
with respect to the PV $\le$~0.1~mm and daughter transverse momentum with respect to the direction
of flight of the mother particle $\ge$~10~MeV/$c$. 
In this plot we see mostly a $\sim 4$~\% statistical error for the
ratios. Extrapolating to the expected $10^8$ events gives an error
of 1.3~\% which should be good enough to decide
between new and old models (see Figure~\ref{f:eta}).

The $D$-meson selection is similar to that of the strange particles using the
decays $D^0 \rightarrow K^- \pi^+$, $\bar{D}^0 \rightarrow K^+ \pi^-$ and $D^\pm
\rightarrow K^{\mp} \pi^{\pm} \pi^{\pm}$. We use a cuts based selection and a
multivariate analysis (MVA)~\cite{ACAT}. MVA is just a more sophisticated method
to cut on the phase space and thus still consistent with being conservative.
Only geometric and kinematic
variables (no significances and no PID) are used for the cuts and in the
MVA. The Minimal requirements for these measurements are a well aligned VeLo and
main tracker. 

In Figure~\ref{f:D0mass} the mass peaks for the $D^0$ and $D^+$ are shown. 
The MVA parameters have been
chosen in a way to get the same signal yield as for the cuts based analysis. One
can see by comparing the plots for the $D^0$ mass peaks that the MVA 
reduces the background by about a factor of three here. For the charged
$D$-mesons we used the cuts based analysis only. The expected sensitivity on $D$
selection is about 2000 particles for each of the above charm species for 100 M
events.  
For $p_t < 12$
GeV, rapidity $1.8 < y < 4.5$ we expect an error on
$\overline{D^0}$/$D^0$ cuts based of 7~\%, MVA 5~\%, for the
$D^-$/$D^+$ we expect to get 6~\%.


\section[Conclusion]{Conclusion and outlook}

With the first $10^{8}$ minimum bias events which corresponds to about one day
of running we expect to measure charged track ratio distributions with $\sim$ 1~\%
error and we expect to be able to probe fragmentation models by strange particle ratios. We expect to reconstruct $\sim$~2000
$D^{0/\pm}$ and the open charm ratios with about 5~\% precision.

More detailed Monte Carlo studies are on the way. In addition we will look into
other strange baryons
($\Xi^-$, $\Omega^-$) ratios and we will look for $b$-baryons like $\Lambda_b$
or $\Xi_b$. Of course corresponding cross section measurements are also planed.


 

\begin{footnotesize}


\bibliographystyle{unsrt}
\bibliography{britsch_markward}
%


\end{footnotesize}


\end{document}